\documentclass[conference]{IEEEtran}
\IEEEoverridecommandlockouts
\usepackage{cite}
\usepackage{amsmath,amssymb,amsfonts}
\usepackage{algorithmic}
\usepackage{graphicx}
\usepackage{textcomp}
\usepackage{xcolor}
\usepackage{hyperref}
\hypersetup{hidelinks,
	colorlinks=true,
	allcolors=black,
	pdfstartview=Fit,
	breaklinks=true}
\def\BibTeX{{\rm B\kern-.05em{\sc i\kern-.025em b}\kern-.08em
    T\kern-.1667em\lower.7ex\hbox{E}\kern-.125emX}}
\begin{document}

\title{MALUNet: A Multi-Attention and Light-weight UNet for Skin Lesion Segmentation\\

\thanks{* Suncheng Xiang and Yuzhuo Fu are the co-corresponding authors. This work was partially supported by the National Natural Science Foundation of China (Grant No. 61977045).}
}

\author{\IEEEauthorblockN{1\textsuperscript{st} Jiacheng Ruan}
\IEEEauthorblockA{\textit{Shanghai Jiao Tong University}\\
Shanghai, China \\
jackchenruan@sjtu.edu.cn}
\and
\IEEEauthorblockN{2\textsuperscript{nd} Suncheng Xiang*}
\IEEEauthorblockA{\textit{Shanghai Jiao Tong University}\\
Shanghai, China \\
xiangsuncheng17@sjtu.edu.cn}
\and
\IEEEauthorblockN{3\textsuperscript{rd} Mingye Xie}
\IEEEauthorblockA{\textit{Shanghai Jiao Tong University}\\
Shanghai, China \\
xiemingye@sjtu.edu.cn}
\and
\IEEEauthorblockN{4\textsuperscript{rd} Ting Liu}
\IEEEauthorblockA{\textit{Shanghai Jiao Tong University}\\
Shanghai, China \\
louisa\_liu@sjtu.edu.cn}
\and
\IEEEauthorblockN{5\textsuperscript{th} Yuzhuo Fu*}
\IEEEauthorblockA{\textit{Shanghai Jiao Tong University}\\
Shanghai, China \\
yzfu@sjtu.edu.cn}
}

\maketitle

\begin{abstract}
Recently, some pioneering works have preferred applying more complex modules to improve segmentation performances. However, it is not friendly for actual clinical environments due to limited computing resources. To address this challenge, we propose a light-weight model to achieve competitive performances for skin lesion segmentation at the lowest cost of parameters and computational complexity so far. Briefly, we propose four modules: (1) \emph{DGA} consists of dilated convolution and gated attention mechanisms to extract global and local feature information; (2) \emph{IEA}, which is based on external attention to characterize the overall datasets and enhance the connection between samples; (3) \emph{CAB} is composed of 1D convolution and fully connected layers to perform a global and local fusion of multi-stage features to generate attention maps at channel axis; (4) \emph{SAB}, which operates on multi-stage features by a shared 2D convolution to generate attention maps at spatial axis. We combine four modules with our U-shape architecture and obtain a light-weight medical image segmentation model dubbed as \emph{MALUNet}. Compared with UNet, our model improves the mIoU and DSC metrics by 2.39\% and 1.49\%, respectively, with a 44x and 166x reduction in the number of parameters and computational complexity. In addition, we conduct comparison experiments on two skin lesion segmentation datasets (ISIC2017 and ISIC2018). Experimental results show that our model achieves state-of-the-art in balancing the number of parameters, computational complexity and segmentation performances. Code is available at \emph{\href{https://github.com/JCruan519/MALUNet}{https://github.com/JCruan519/MALUNet}}.
\end{abstract}


\section{Introduction}
Medical image segmentation models can assist doctors in diagnosis and improve the efficiency of hospital operations. Most existing methods are based on UNet\cite{unet}, an Encoder-Decoder model with an U-shape architecture. Due to its simplicity and scalability advantages, plenty of improved models are proposed, such as UNet++\cite{unet++}, AttentionUNet\cite{attentionunet}, 3D-UNet\cite{3DUNet}, V-Net\cite{vnet}, UNet3+\cite{unet3+} and so on. Recently, researchers have introduced the self-attention mechanism (SA) of ViT\cite{vit} into UNet. Some essential extensions like TransUNet\cite{transunet}, TransBTS\cite{transbts}, Swin-UNet\cite{swinunet}, DS-TransUNet\cite{dstransunet}, TransFuse\cite{transfuse}, UTNet\cite{utnet} and UTNetV2\cite{utnetv2} have been proposed, further improving segmentation performances.

\begin{figure}[!t]
	\centerline{\includegraphics[width=9cm, height=4.5cm]{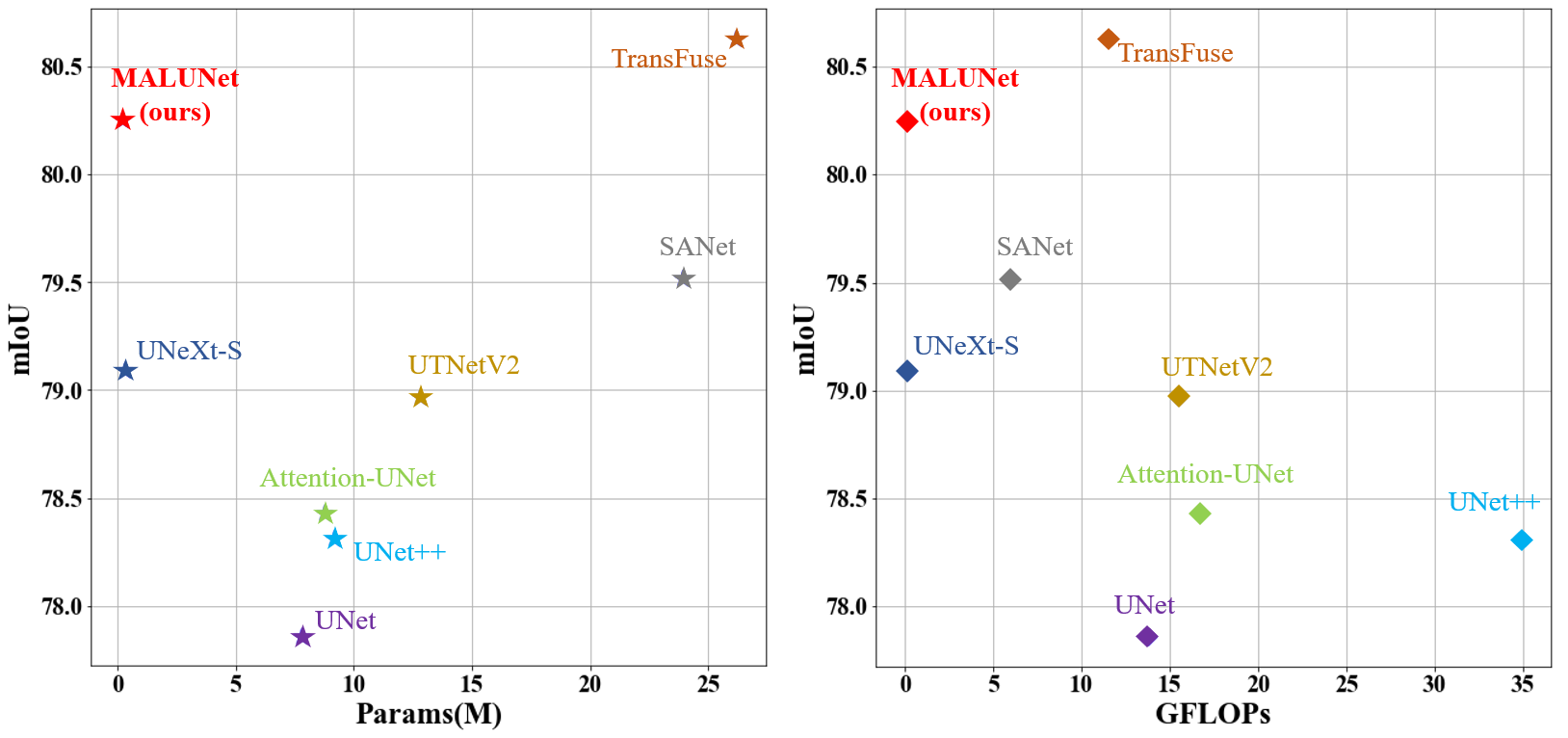}}
	\caption{Visualization of comparison results on the ISIC2018 dataset. X-axis corresponds to the number of parameters and GFLOPs, respectively (lower the better). Y-axis represents the mIoU (higher the better).}
	\label{fig1}
\end{figure}

However, the previous works still have the following problems. Firstly, previous studies tend to introduce more complex modules into UNet in exchange for improving performances. However, due to the limited memory of mobile medical devices, many models with a large amount of parameters cannot be employed well in real-world clinical scenarios. Secondly, medical image segmentation belongs to the layout-specific task\cite{tvconv}. Namely, the variance between samples is slight, but within samples is significant in the medical dataset. The modules in recent works are often improved based on the SA module or its variants, while SA is not expert in modeling the relationship between samples\cite{EAmlp}. Finally, for medical image segmentation, it is imperative to fully use multi-stage and multi-scale information due to varying target sizes. However, many models ignore the importance of that.

In this paper, we propose four modules to solve the problems: (1) Dilated Gated Attention Block (DGA), which helps our model pay more attention to the target region by obtaining global and local information. At the same time, the depthwise separable convolution\cite{mobilenets} further reduces the number of parameters and computational complexity. (2) An efficient external attention mechanism for the light-weight medical segmentation model is presented, called Inverted External Attention Block (IEA), which can enhance the information association between samples and obtain characteristics of the overall dataset. (3) For the acquisition of multi-stage and multi-scale information, attention bridge modules are proposed for channel and spatial axes, dubbed as Channel Attention Bridge Block (CAB) and Spatial Attention Bridge Block (SAB), which can generate corresponding attention maps. With the help of the modules mentioned above, we could drastically reduce the number of channels in our model and maintain a competitive segmentation effect with the minimum number of parameters and lowest complexity.

The contributions lie in three folds: 
\begin{itemize}
\item Four attention modules are proposed, which could obtain global and local feature information respectively, depict the sample characteristics of the whole dataset, and fuse multi-stage information to generate corresponding attention maps at the channel and spatial levels. 
\item We present MALUNet based on our U-shape architecture and four attention modules, which has the minimum amount of parameters and the lowest computational complexity. 
\item Extensive experiments are performed on ISIC2017\cite{isic17} and ISIC2018\cite{isic18} datasets. Results show that our model achieves the state-of-the-art in terms of the balance between parameters, computational complexity and segmentation performances.
\end{itemize}

\section{Related Works}
FCN\cite{fcn} is the pioneer in image segmentation, which utilizes fully convolution to extract feature information. After that, UNet\cite{unet}, an Encoder-Decoder model based on the fully convolution, is developed for medical image segmentation and achieves terrific results.

Most medical image segmentation models are improved based on UNet, such as UNet++\cite{unet++}, Attention-UNet\cite{attentionunet}, Res-UNet\cite{resunet}\cite{resunet1}, Dense-UNet\cite{denseunet}, etc. UNet++ replaces the cropping and concatenating operation in the skip-connection part of UNet with the convolution operation in a dense way to obtain better feature information and compensate for the information loss caused by sampling. Attention-UNet utilizes attention gates to control the importance of features, making it pay more attention to targets. In order to further decrease the loss of information and improve performances, Res-UNet and Dense-UNet replace plain convolutions with Res-block in ResNet\cite{resnet} and Dense-block in DenseNet\cite{densenet}, respectively.

Google transplanted the SA in natural language processing\cite{transformer} to computer vision, and proposed ViT\cite{vit} as the backbone. Because of its powerful feature extraction ability, researchers recently focused on how to combine ViT and its variants with UNet to obtain better results\cite{trmmedicalsurvey}\cite{trmmedicalreview}. For example, Swin-UNet\cite{swinunet} combines Swin Transformer\cite{swintrm} with UNet and receives a better segmentation effect. DS-TransUNet\cite{dstransunet} utilizes patches of different scales as input, and two parallel Swin Transformers are used as encoders to attain more abundant information. TransFuse\cite{transfuse} applies the parallel encoder structure of CNN and Transformer to obtain local and global information simultaneously. UTNetV2\cite{utnetv2} improves the UTNet\cite{utnet}, which is a hybrid multi-level structure. Local modeling is introduced to reduce the dependence on large-scale data via using depthwise separable convolution as the projection and feed-forward network in the Transformer block.

The above networks are designed to improve performances while ignoring the shortcomings of heavy parameters and computational complexity, making it difficult to be applied in the real medical environment. Recently, Jeya et al. proposed UNeXt\cite{unext} based on combining MLP\cite{mlp} with UNet, which significantly reduces the number of parameters on the premise of ensuring the segmentation performances. A light-weight model is necessary for practical applications and mobile health. Therefore, based on previous studies, this paper proposes a light-weight model for medical image segmentation, and introduces a variety of attention modules to ensure performances.

\begin{figure}[!t]
	\centerline{\includegraphics[width=9cm, height=8cm]{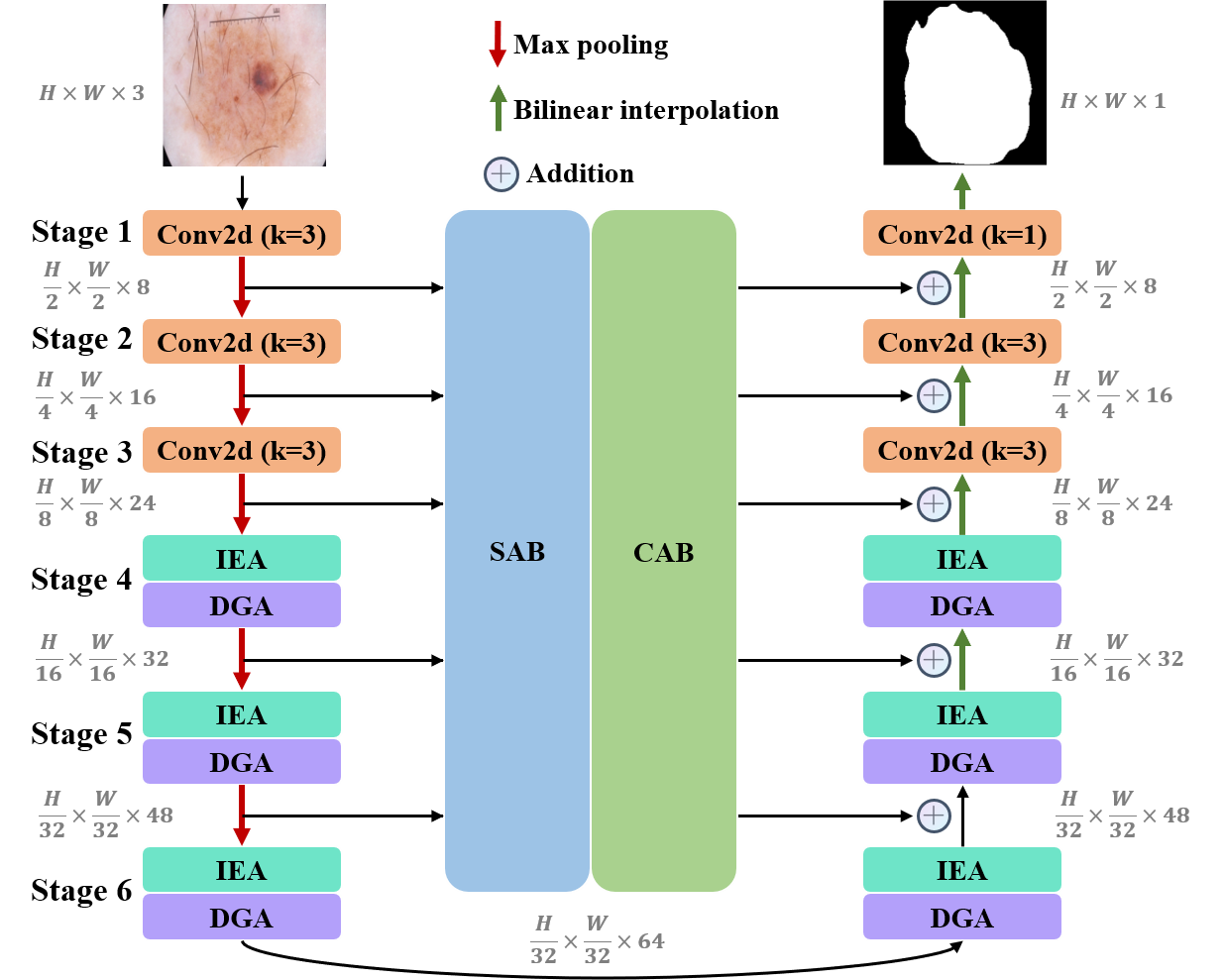}}
	\caption{The illustration of MALUNet architecture.}
	\label{fig2}
\end{figure}

\section{Methods}
In this section, we first introduce the four modules proposed in this paper: DGA, IEA, CAB and SAB. Afterward, we elaborate on the proposed light-weight model, MALUNet, as shown in Fig. \ref{fig2}.

\subsection{Dilated Gated Attention Block}
Medical image segmentation belongs to dense prediction tasks, and the simultaneous acquisition of global and local information is crucial to improving performances. Global information is helpful for the model to understand the overall lesion structure and its relationship with the background, making it locate the lesion area more accurately. Moreover, local information helps to obtain the details of edges and corners of the lesion area, making predictions more complete.

\begin{figure}[!t]
	\centerline{\includegraphics[width=9cm, height=3.6cm]{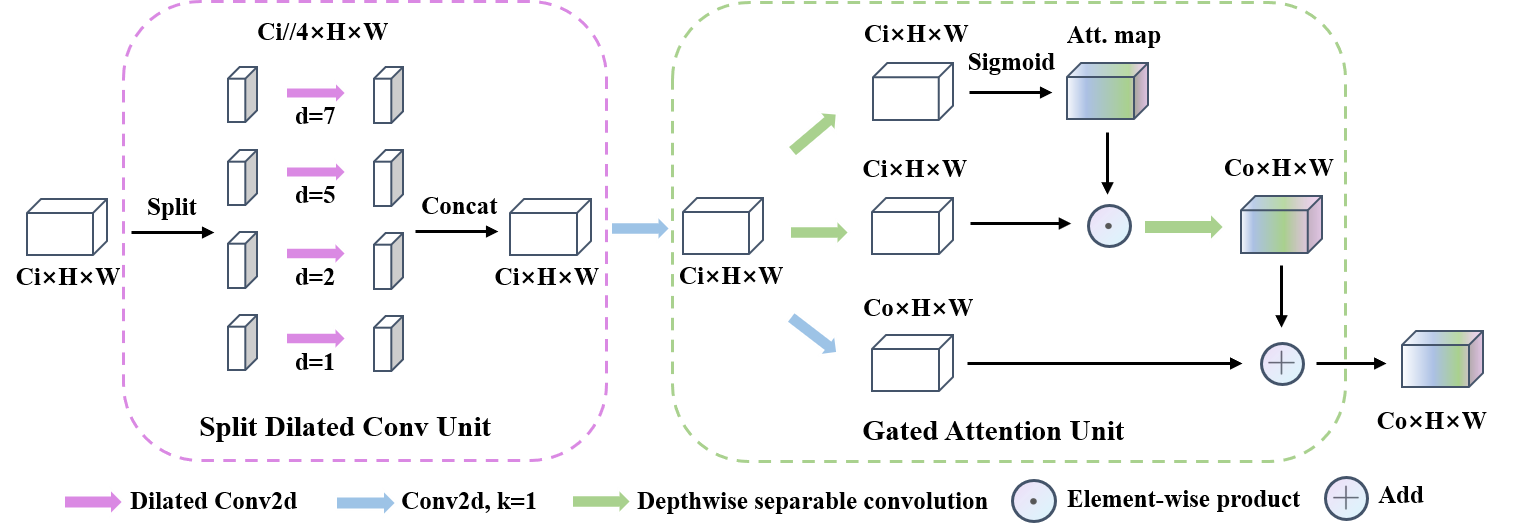}}
	\caption{Overview of the proposed DGA architecture. Note that unless otherwise specified, the convolution kernel size is 3 × 3.}
	\label{fig3}
\end{figure}

Therefore, DGA is proposed in this paper. As shown in Fig. \ref{fig3}, this module is composed of two sub-units: Split Dilated Conv Unit (SDC) and Gated Attention Unit (GA). SDC splits the feature map along the channel dimension into four parts, and obtains the global (convolutions with dilated rates of 5 and 7) and local (convolutions with dilated rates of 1 and 2) feature information through depthwise separable convolution with different dilated rates\cite{mobilenets}\cite{dilatedconvolutions}. Then, the concatenating operation is carried out in the channel dimension to restore the size of the feature map, followed by a convolution operation to interact with the global and local information. Next, for GA, the attention map with the same shape as the input feature is generated via depthwise separable convolution to suppress the unimportant region of feature information transmitted by SDC, so that model pays more attention to the vital information. Finally, a residual connection operation is applied to obtain the output. The above process can be expressed by formulas (1) to (5).

\begin{equation}
x_{1}, x_{2}, x_{3}, x_{4}=\operatorname{Chunk_{4}}(X)
\end{equation}
\begin{equation}
x_{1}^{\prime}, x_{2}^{\prime}, x_{3}^{\prime}, x_{4}^{\prime}=\operatorname{W_{1}}\left(x_{1}\right), \operatorname{W_{2}}\left(x_{2}\right), \operatorname{W_{5}}\left(x_{3}\right), \operatorname{W_{7}}\left(x_{4}\right)
\end{equation}
\begin{equation}
X^{\prime}=\operatorname{W}\left(\operatorname{Concat}\left(x_{1}^{\prime}, x_{2}^{\prime}, x_{3}^{\prime}, x_{4}^{\prime}\right)\right)
\end{equation}
\begin{equation}
Att=\sigma\left(\operatorname{DW}\left(X^{\prime}\right)\right)
\end{equation}
\begin{equation}
Out=\operatorname{DW}\left(\operatorname{DW}\left(X^{\prime}\right) \odot A t t\right)+\operatorname{W}\left(X^{\prime}\right)
\end{equation}

Where $\operatorname{Chunk_{4}}$ indicates that the input feature map is divided into four parts along the channel dimension, $\operatorname{W_{i}}$ represents depthwise separable convolution with dilated rates of $i$, $\operatorname{Concat}$ denotes the concatenating operation, $\operatorname{W}$ represents the plain convolution operation, $\sigma$ is the sigmoid function, $\operatorname{DW}$ indicates the depthwise separable convolution, and $\odot$ represents the element-wise multiplication.

\subsection{Inverted External Attention Block}
\begin{figure}[!t]
	\centerline{\includegraphics[width=9cm, height=3.6cm]{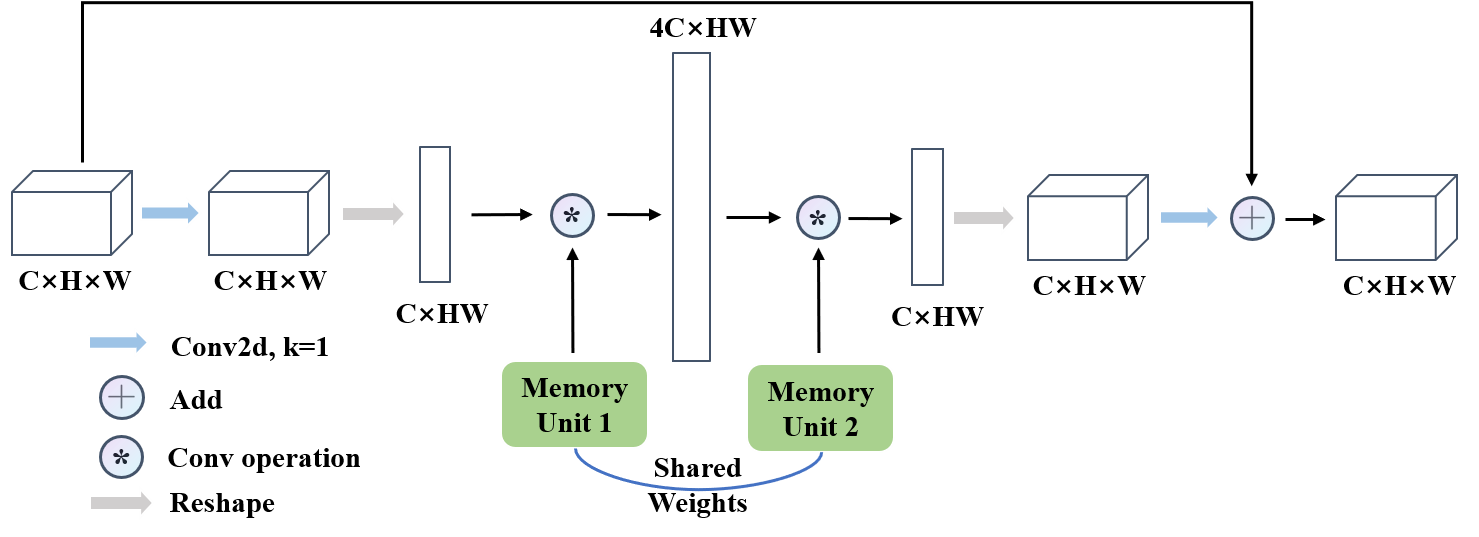}}
	\caption{The illustration of IEA architecture.}
	\label{fig4}
\end{figure}

EAMLP\cite{EAmlp} proposes a new attention mechanism, called external attention (EA), which employs two memory units (two Conv1D operations that share parameters) to characterize the feature information of the entire dataset. Besides, inspired by the inverted residual module in MobileNetV2\cite{mobilenetv2}, a general light-weight backbone in computer vision, we propose an inverted external attention block (IEA) for light-weight medical image segmentation model as shown in Fig. \ref{fig4}.

Given an input $X\in\mathcal{R}^{C \times H \times W}$, after 1 × 1 convolution and reshaping operation, the size is changed to $X\in \mathcal{R}^{C \times HW}$. We use Memory Unit 1 to expand the feature map fourfold to obtain $Att\in \mathcal{R}^{4C \times HW}$, and then Memory Unit 2 is applied to restore the dimension, followed by reshaping operation to restore to the original feature map size. Finally, the output feature map can be obtained through a 1 × 1 convolution and added with the residual information. It is worth noting that in IEA, we no longer fix the number of channels between two memory units as 64 as EA, but apply four times expansion. The advantage is that the memory unit maps the input to a higher dimensional space, which makes the memory unit describe the overall feature information of the dataset more comprehensively.

\subsection{Channel Attention Bridge Block}
The acquisition of multi-stage and multi-scale information plays an essential role in segmenting targets of different sizes, and the fusion of multi-stage and multi-scale information has been proved to be the key to improve the performance. Therefore, we propose a bridge attention module for channel level, called CAB. It is utilized to generate channel attention map by concatenating the features of different stages at channel axis to better integrate information. CAB can be expressed by formulas (6) to (10).

\begin{equation}
t_{i}^{\prime}=\operatorname{GAP}\left(t_{i}\right)
\end{equation}
\begin{equation}
T=\operatorname{Concat}\left(t_{1}^{\prime}, t_{2}^{\prime}, \ldots, t_{s-1}^{\prime}\right)
\end{equation}
\begin{equation}
T^{\prime}=\operatorname{Conv1D}(T)
\end{equation}
\begin{equation}
A t t_{i}=\sigma\left(\operatorname{FC}_{i}\left(T^{\prime}\right)\right)
\end{equation}
\begin{equation}
Out_{i}=t_{i}+t_{i} \odot Att_{i}
\end{equation}

Where $\operatorname{GAP}$ refers to global average pooling, $t_i$ represents the feature map of different stages obtained from the encoder, $\operatorname{Concat}$ indicates concatenating operation in the channel dimension, $s$ denotes the total number of stages, $\operatorname{Conv1D}$ represents 1D convolution operation, $\operatorname{FC}_{i}$ is the fully connected layer for stage $i$, $\sigma$ is the sigmoid function, and $\odot$ is the element-wise multiplication.

\begin{figure}[!t]
	\centerline{\includegraphics[width=9.5cm, height=4.5cm]{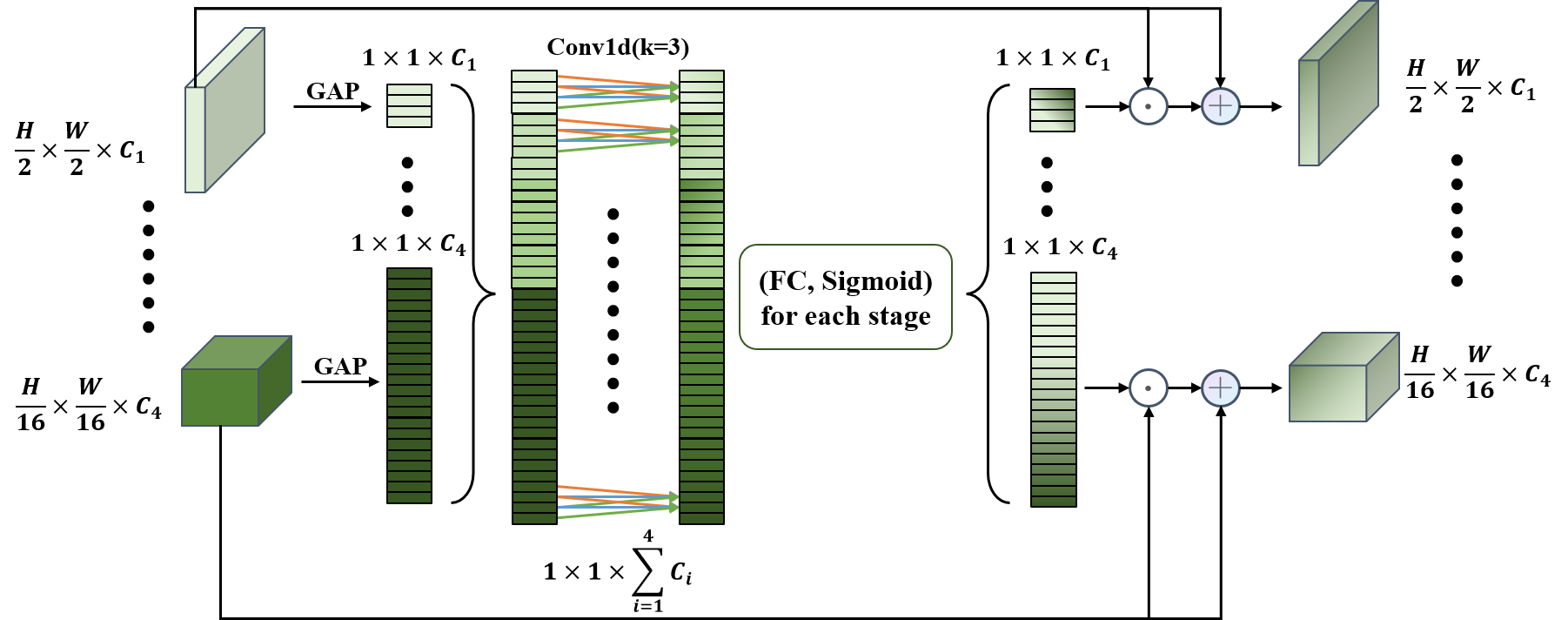}}
	\caption{Overview of the proposed CAB architecture.}
	\label{fig5}
\end{figure}

As shown in Fig. \ref{fig5}, CAB is visualized by taking the five-stage model ($s=5$) as an example. Note that CAB subdivides the multi-stage and multi-scale information fusion into local information fusion (1D convolution operations) and global information fusion (different fully connected layers for each stage), to provide a more informative attention feature map.

\subsection{Spatial Attention Bridge Block}
\begin{figure}[!t]
	\centerline{\includegraphics[width=9cm, height=4cm]{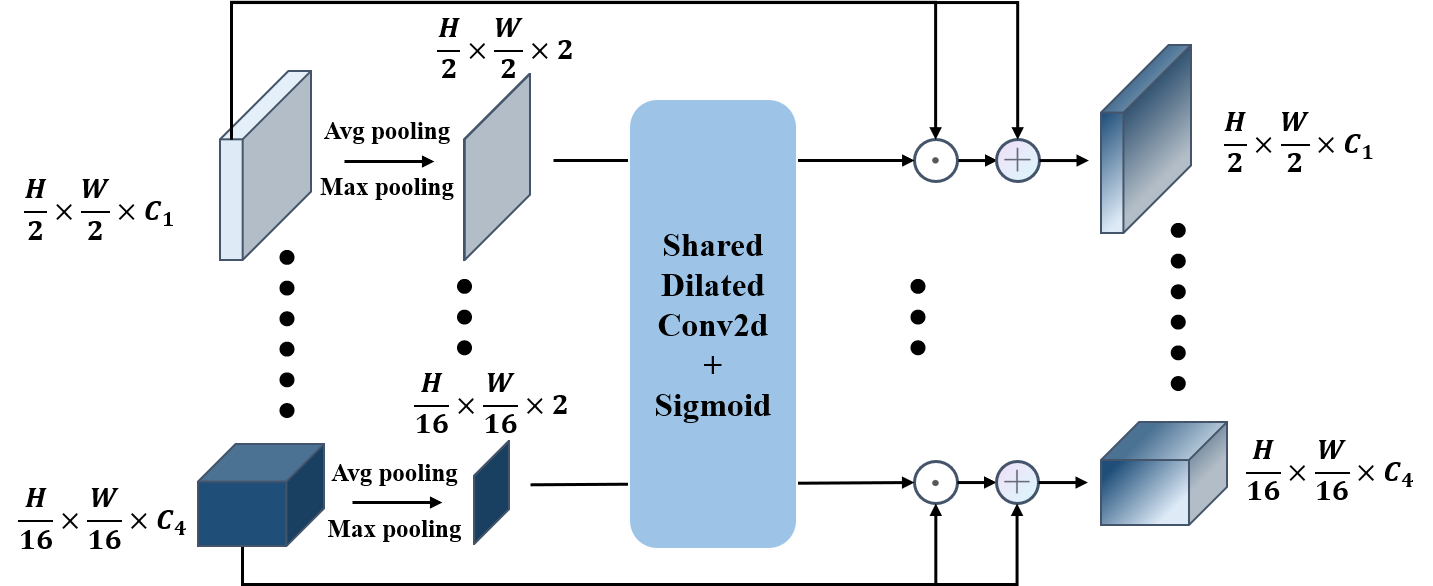}}
	\caption{Overview of the proposed SAB architecture.}
	\label{fig6}
\end{figure}

CBAM\cite{cbam} believes that both channel and spatial attention play an essential role in suppressing irrelevant information. Therefore, inspired by CBAM, we introduced another bridge attention module, SAB, as shown in Fig. \ref{fig6}. Also, five-stage SAB is visualized as an example. This module fuses the multi-stage and multi-scale information at spatial axis to generate the attention map for each stage.

First, for each stage's feature map, we employ average-pooling and max-pooling operations in the channel dimension, and concatenate them to get the feature map with two channels, while the height and width remain unchanged. Then, a shared dilated convolution operation is utilized (the dilated rate is 3 and kernel size is 7), followed by a sigmoid function to generate a spatial attention map for each stage. Finally, the generated spatial attention map is element-wise multiplied by the original map and added with the residual information.

\subsection{MALUNet}
After combining our attention modules and U-shape architecture, our model is obtained, dubbed as MALUNet, as shown in Fig. \ref{fig2}.

In MALUNet, a six-stage U-shape architecture is applied, and the number of channels in each stage is \{8, 16, 24, 32, 48, 64\}. Based on UNet, we add one stage, and the plain convolution operation is reduced from two to one in Stage 1-3. In Stage 4-6, we use IEA and DGA, which are arranged in series. The former depicts the overall features of the dataset and the relationship between samples, and the latter acquires the global and local feature information in samples. They complement each other and make full use of the advantages of different attention mechanisms. In addition, SAB-first order is employed for the sequential arrangement of two bridge attention modules. The two bridge attention modules can fuse the multi-stage and multi-scale features of Stages 1-5 to generate the attention maps in the spatial and channel dimension. And then, we add features obtained by bridge attention modules with features of the decoder part to reduce the feature semantic difference between the encoder and decoder while alleviating the information loss caused by the sampling process.

\section{Experiments}

\subsection{Datasets}
In this section, we conduct extensive experiments on two public datasets of skin lesion segmentation, the International Skin Imaging Collaboration challenge dataset (ISIC2017 and ISIC2018)\cite{isic17}\cite{isic18}, to train and evaluate the proposed model. ISIC2017 and ISIC2018 have 2150 and 2694 dermoscopy images with segmentation mask labels, respectively. We randomly divide datasets in a ratio of 7 : 3 as our experiments' training and testing sets. For ISIC2017, there are 1500 images in training sets and 650 images in testing sets. For ISIC2018, there are 1886 images in training sets and 808 images in testing sets. Note that we perform comparative experiments on ISIC2017 and ISIC2018 and conduct ablation experiments on ISIC2018.

\subsection{Implementation details}
All experiments are implemented on a single NVIDIA GeForce RTX3080 GPU. Empirically, all images are normalized and resized to 256 × 256, and we employ data augmentation including vertical flip, horizontal flip and random rotation. The loss function is BceDice loss, which can be expressed by formula (11). We utilize AdamW\cite{adamw} as the optimizer with an initial learning rate of 0.001. A cosine annealing learning rate scheduler is used with a maximum number of iterations of 50 and a minimum learning rate of 0.00001. The training epoch is set equal to 300, and the batch size is 8. 

\begin{equation}
	\begin{gathered}
		L_{Bce}=-\frac{1}{N} \sum_{i=1}^{N}\left[y_{i} \log \left(p_{i}\right)+\left(1-y_{i}\right) \log \left(1-p_{i}\right)\right] \\
		L_{Dice}=1-\frac{2|X \cap Y|}{|X|+|Y|} \\
		L_{BceDice}=\lambda_{1} L_{Bce}+\lambda_{2} L_{Dice}
	\end{gathered}
\end{equation}

Where $N$ is the total number of samples, $y_i$ is the real label, $p_i$ is the prediction. $|X|$ and $|Y|$ represent ground truth and prediction, respectively. $\lambda_1$ and $\lambda_2$ refer to the weight of two loss functions. In this paper, both weights are taken as 1 by default.

\subsection{Evaluation Metrics}
Five metrics including Mean Intersection over Union (\textbf{mIoU}), Dice similarity score (\textbf{DSC}), Accuracy (\textbf{Acc}), Sensitivity (\textbf{Sen}), Specificity (\textbf{Spe}) are used to measure segmentation performances. In addition, \textbf{Params} is utilized to indicate the number of parameters, and the unit is Million (M). The computational complexity is calculated regarding the number of floating point operators (\textbf{GFLOPs}). Note that the \textbf{Params} and \textbf{GFLOPs} of models are measured with 256 × 256 input size.

\begin{equation}
	\begin{cases}
		\text{\textbf{mIoU}}=\frac{TP}{TP+FP+FN} \\
		\text{\textbf{DSC}}=\frac{2TP}{2TP+FP+FN} \\
		\text{\textbf{Acc}}=\frac{TP+TN}{TP+TN+FP+FN} \\
		\text{\textbf{Sen}}=\frac{TP}{TP+FN} \\
		\text{\textbf{Spe}}=\frac{TN}{TN+FP} \\
	\end{cases}
\end{equation}

Where $TP$, $FP$, $FN$, $TN$ represent true positive, false positive, false negative, and true negative.

\subsection{Comparison with state-of-the-arts}
This section compares MALUNet with state-of-the-arts that employ different ideas to improve UNet\cite{unet} in recent years on ISIC2017 and ISIC2018 datasets. The experimental results are shown in Table \ref{tab1} and Table \ref{tab2}. Note that UNeXt\cite{unext}, a light-weight medical image segmentation model, is based on the five-stage U-shape architecture, and the number of channels in each stage is \{32, 64, 128, 160, 256\} in the original paper. In order to compare more obviously with our model, we employ UNeXt-S, and its number of channels is \{8, 16, 32, 64, 128\} in our experiments, which is the default setting used in its official open source codes. Therefore, the parameters and computational complexity of UNeXt-S and our model are the same sizes, which is convenient to compare the performance. 

\begin{figure*}[!t]
	\centerline{\includegraphics[width=18cm, height=6.5cm]{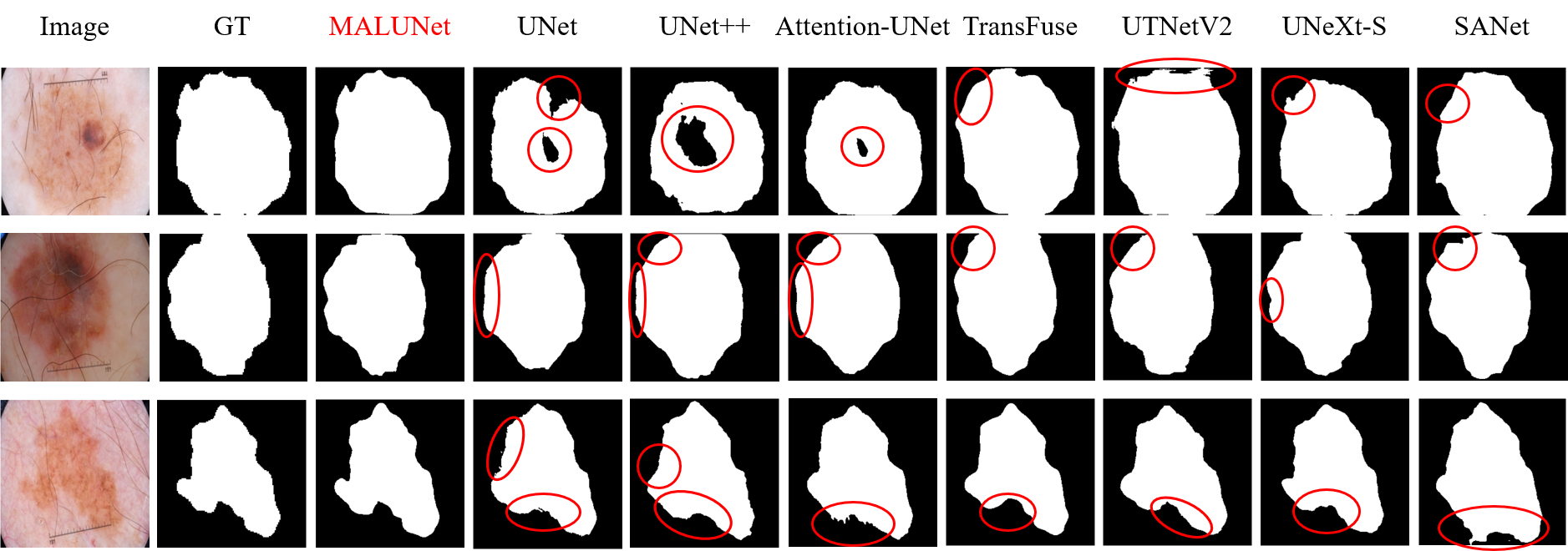}}
	\caption{Visualization of some predictions. The red mark circles defect parts of the predictions.}
	\label{fig7}
\end{figure*}

According to the results, compared with UNet, MALUNet still improves the performance in an all-around way. In addition, the parameters are reduced by 44x, and the computational complexity is decreased by 166x. For ISIC2018 dataset, compared with TransFuse\cite{transfuse}, although our model is 0.38\% and 0.23\% lower in mIoU and DSC, it can not be ignored that MALUNet is 150x and 139x lower than TransFuse in terms of parameters and computational complexity. Compared with UNeXt-S in the same starting point as ours, the performance of MALUNet on the two datasets is better than UNeXt-S, and further reduces the number of parameters and computational complexity by 0.125M and 0.017GFLOPs. As shown in Fig. \ref{fig1}, it can be seen more clearly that MALUNet achieves state-of-the-art in terms of the balance between parameters, computational complexity and segmentation performances. Besides, we visualize some results, as shown in Fig. \ref{fig7}, and it is evident that our model achieves better performances with more precise edges and no hollows.

\begin{table}[!t]
	\setlength\tabcolsep{3pt}
	\renewcommand\arraystretch{1.25}
	\scriptsize
	\caption{Comparative experimental results on ISIC2017 dataset. (\textbf{bold} indicates the best and \underline{underline} indicates the second best.)}
	\begin{center}
		\begin{tabular}{c|cc|ccccc}
			\hline
			\textbf{Model} & \textbf{Params} $\downarrow$ & \textbf{GFLOPs}$\downarrow$ & \textbf{mIoU}$\uparrow$  & \textbf{DSC}$\uparrow$   & \textbf{Acc}$\uparrow$   & \textbf{Spe}$\uparrow$   & \textbf{Sen}$\uparrow$   \\ \hline
			UNet\cite{unet}    & 7.77       & 13.78      & 76.98 & 86.99 & 95.65 & 97.43       & 86.82       \\
			TransFuse\cite{transfuse}      & 26.27           & 11.53           & \textbf{79.21} & \textbf{88.40} & \underline{96.17}    & 97.98          & \textbf{87.14} \\
			UTNetV2\cite{utnetv2} & 12.80      & 15.50      & 77.35 & 87.23 & 95.84 & \underline{98.05} & 84.85       \\
			UNeXt-S\cite{unext}   & \underline{0.30} & \underline{0.10} & 78.26 & 87.80 & 95.95 & 97.74       & \underline{87.04} \\
			MALUNet (ours)  & \textbf{0.175}  & \textbf{0.083}  & \underline{78.78}    & \underline{88.13}    & \textbf{96.18} & \textbf{98.47} & 84.78          \\ \hline
		\end{tabular}
		\label{tab1}
	\end{center}
\end{table}

\begin{table}[!t]
	\setlength\tabcolsep{3pt}
	\renewcommand\arraystretch{1.25}
	\scriptsize
	\caption{Comparative experimental results on ISIC2018 dataset. (\textbf{bold} indicates the best and \underline{underline} indicates the second best.)}
	\begin{center}
		\begin{tabular}{c|cc|ccccc}
			\hline
			\textbf{Model} & \textbf{Params} $\downarrow$ & \textbf{GFLOPs}$\downarrow$ & \textbf{mIoU}$\uparrow$  & \textbf{DSC}$\uparrow$   & \textbf{Acc}$\uparrow$   & \textbf{Spe}$\uparrow$   & \textbf{Sen}$\uparrow$  \\ \hline
			UNet\cite{unet}           & 7.77       & 13.78      & 77.86 & 87.55 & 94.05 & \underline{96.69}    & 85.86 \\
			UNet++\cite{unet++}         & 9.16       & 34.90      & 78.31 & 87.83 & 94.02 & 95.75          & 88.65 \\
			Attention-UNet\cite{attentionunet} & 8.73       & 16.74      & 78.43 & 87.91 & 94.13 & 96.23          & 87.60 \\
			TransFuse\cite{transfuse}      & 26.27           & 11.53           & \textbf{80.63} & \textbf{89.27} & \textbf{94.66} & 95.74        & \textbf{91.28} \\
			UTNetV2\cite{utnetv2}        & 12.80      & 15.50      & 78.97 & 88.25 & 94.32 & 96.48          & 87.60 \\
			SANet\cite{sanet}          & 23.90      & 5.99       & 79.52 & 88.59 & 94.39 & 95.97          & 89.46 \\
			UNeXt-S\cite{unext}          & \underline{0.30} & \underline{0.10} & 79.09 & 88.33 & 94.39 & \textbf{96.72} & 87.15 \\
			MALUNet (ours)  & \textbf{0.175}  & \textbf{0.083}  & \underline{80.25}    & \underline{89.04}    & \underline{94.62}    & 96.19        & \underline{89.74}    \\ \hline
		\end{tabular}
		\label{tab2}
	\end{center}
\end{table}

\subsection{Ablation Studies}
\paragraph{Determination of baseline}
UNet\cite{unet} is a five-stage U-shape architecture model. The number of channels in each stage is \{32, 64, 128, 256, 512\}, and the number of plain convolutions in each stage is two. In this paper, we change the five-stage U-shape architecture to six stages, and the number of plain convolutions in each stage is one. Table \ref{tab3} shows the effect of baseline under different settings. It is noted that the UNet in five-stage uses concatenating operation between encoder and decoder, and the six-stage model is element-wise addition. Under the premise of considering the performance and the size of model comprehensively, we utilize the six-stage UNet with the number of channels of \{8, 16, 24, 32, 48, 64\}, dubbed as BasicUNet (BU), to carry out the subsequent ablation experiments.

\begin{table}[!t]
	\setlength\tabcolsep{1.2pt}
	\renewcommand\arraystretch{1.25}
	\scriptsize
	\caption{Ablation performances of UNet under different settings.}
	\begin{center}
		\begin{tabular}{c|cc|ccccc}
			\hline
			\textbf{\#Channels}       & \textbf{Params} $\downarrow$ & \textbf{GFLOPs}$\downarrow$ & \textbf{mIoU}$\uparrow$  & \textbf{DSC}$\uparrow$   & \textbf{Acc}$\uparrow$   & \textbf{Spe}$\uparrow$   & \textbf{Sen}$\uparrow$   \\ \hline
			\{32,64,128,256,512\}   & 7.77            & 13.76           & 77.86          & 87.55          & 94.05          & 96.69 & 85.86          \\
			\{16,32,64,128,160,256\} & 1.30            & 0.42            & \textbf{78.59} & \textbf{88.01} & 94.15          & 96.07          & \textbf{88.16} \\
			\{8,16,32,64,128,160\}   & 0.57            & 0.14            & 78.50          & 87.96          & \textbf{94.24} & \textbf{96.75}          & 86.42          \\
			\{8,16,32,48,64,96\}     & 0.21            & 0.10            & 78.44          & 87.92          & 94.11          & 96.08          & 87.98          \\
			\{8,16,24,32,48,64\}     & \textbf{0.11}   & \textbf{0.07}   & 78.22          & 87.77          & 94.09          & 96.32          & 87.15          \\ \hline
		\end{tabular}
		\label{tab3}
	\end{center}
\end{table}

\paragraph{Ablation study on the single module}
The ablation experiment of the single module is based on BasicUNet. We train and test each proposed module on ISIC2018 dataset to prove its effectiveness. Moreover, to illustrate the advantages of our IEA, we also employ EA\cite{EAmlp} for comparison. Note that (1) the number of intermediate channels in EA used here is the same as the original paper, which is set to 64; (2) DGA replaces the convolution operation in BasicUNet in Stage 4-6; (3) EA and IEA are utilized in Stage 4-6 before the plain convolution; (4) CAB and SAB fuse features from the encoder of Stage 1-5. As shown in Table \ref{tab4}, DGA can improve the performance while reducing the amount of parameters and computational complexity at the same time. It can be shown that obtaining global and local feature information simultaneously can make the model ``see'' more comprehensively. In addition, the other three modules can also improve the performance while only a few additional parameters are introduced.

\begin{table}[!t]
	\setlength\tabcolsep{2pt}
	\renewcommand\arraystretch{1.25}
	\scriptsize
	\caption{Ablation performances of the single module.}
	\begin{center}
		\begin{tabular}{c|cc|ccccc}
			\hline
			\textbf{Module}   & \textbf{Params} $\downarrow$ & \textbf{GFLOPs}$\downarrow$ & \textbf{mIoU}$\uparrow$  & \textbf{DSC}$\uparrow$   & \textbf{Acc}$\uparrow$   & \textbf{Spe}$\uparrow$   & \textbf{Sen}$\uparrow$   \\ \hline
			BasicUNet (BU)          & 0.11 & 0.07 & 78.22 & 87.77 & 94.09 & 96.32 & 87.15 \\
			BU + DGA & \textbf{0.08}   & \textbf{0.06}   & 79.12          & 88.43          & 94.33          & 96.29          & \textbf{88.25} \\
			BU + EA  & 0.15 & 0.08 & 79.31 & 88.46 & 94.44 & 96.65 & 87.57 \\
			BU + IEA & 0.18            & 0.09            & \textbf{79.77} & \textbf{88.75} & \textbf{94.58} & \textbf{96.77} & 87.76          \\
			BU + CAB & 0.12 & 0.07 & 78.98 & 88.26 & 94.36 & 96.69 & 87.12 \\
			BU + SAB & 0.11 & 0.07 & 78.61 & 88.02 & 94.24 & 96.60 & 86.92 \\ \hline
		\end{tabular}
		\label{tab4}
	\end{center}
\end{table}

\paragraph{Ablation study on four modules}
Four modules hybrid ablation experiment on ISIC2018 is utilized to determine the best form of MALUNet. The ablation experiment results are shown in Table \ref{tab5}. After combining the four modules with BasicUNet (BU), although the performance of BU+DGA+IEA+(B) is the best, the starting point of this paper is to design a light-weight medical image segmentation model by balancing the relationship between the parameters, computational complexity and performances. Therefore, BU+IEA+DGA+(B) is finally selected as MALUNet in this paper.

\begin{table}[!t]
	\setlength\tabcolsep{2pt}
	\renewcommand\arraystretch{1.25}
	\scriptsize
	\caption{Ablation performances of four modules. (A) represents the CAB-first order and (B) represents the SAB-first order.}
	\begin{center}
		\begin{tabular}{c|cc|ccccc}
			\hline
			\textbf{Model} & \textbf{Params} $\downarrow$ & \textbf{GFLOPs}$\downarrow$ & \textbf{mIoU}$\uparrow$  & \textbf{DSC}$\uparrow$   & \textbf{Acc}$\uparrow$   & \textbf{Spe}$\uparrow$   & \textbf{Sen}$\uparrow$ \\ \hline
			BU+DGA+IEA+(A) & 0.247          & 0.097          & \textbf{80.35} & 89.10          & 94.71          & 96.57          & 88.93          \\
			BU+DGA+IEA+(B) & 0.247          & 0.097          & 80.23          & \textbf{89.14} & \textbf{94.84} & \textbf{97.34} & 87.00          \\
			BU+IEA+DGA+(A) & \textbf{0.175} & \textbf{0.083} & 79.80          & 88.77          & 94.42          & 95.65          & \textbf{90.58} \\
			BU+IEA+DGA+(B) & \textbf{0.175} & \textbf{0.083} & 80.25          & 89.04          & 94.62          & 96.19          & 89.74          \\ \hline
		\end{tabular}
		\label{tab5}
	\end{center}
\end{table}

\section{Conclusion}
In this study, we propose four attention modules in order to: 1) Obtain global and local feature information; 2) Depict the sample feature of the whole dataset so that the relationship between samples can be established; 3) Generate a complete channel attention map by fusing the multi-stage and multi-scale features locally and globally at channel axis; 4) Fuse the multi-stage and multi-scale feature information, and obtain the corresponding spatial attention map at spatial axis. MALUNet is presented by combining our proposed modules and six-stage U-shape architecture, which is a light-weight medical image segmentation model. The comparative experiments and visualization results on ISIC2017 and ISIC2018 datasets can prove that we realize state-of-the-art in terms of the balance between parameters, computational complexity and performances. It is believed that our work provides new enlightenment for the subsequent development of light-weight medical image segmentation models. However, the limitation is that MALUNet is proposed only for skin lesion segmentation. Therefore, we will improve performances through pruning and neural architecture search methods and make our model applicable to other diseases in future work.

{\bibliographystyle{IEEEtran}
\bibliography{egbib}}

\end{document}